\documentclass[aps,prd,twocolumn,groupedaddress,showpacs,nofootinbib,amssymb]{revtex4-2}
\usepackage[dvips]{graphicx}
\usepackage{amssymb}
\usepackage{amsmath}
\usepackage{graphicx,,color}
\usepackage{amsfonts}
\usepackage{bm}
\usepackage{cancel}
\usepackage{comment}
\usepackage{braket}

\newcommand\be{\begin{equation}}
\newcommand\ee{\end{equation}}

\newcommand\e{\mathrm{e}}
\DeclareUnicodeCharacter{2212}{-}
\allowdisplaybreaks[4]

\begin{document}

\tolerance=5000

\title{Inflationary phenomenology of non-minimally coupled
\\
Einstein-Chern-Simons gravity}
\author{F.P. Fronimos}\, \email{fotisfronimos@gmail.com}
\author{S.A. Venikoudis}\, \thanks{venikoudis@gmail.com}

\affiliation{
 Department of Physics, Aristotle University of
Thessaloniki, Thessaloniki 54124,
Greece
}

\tolerance=5000

\begin{abstract}
In this work we investigate the inflationary era in the presence of a canonical scalar field and Chern-Simons parity violating corrections. It was also assumed that a non minimal coupling between curvature and the scalar field is present. For the shake of completeness, the slow-roll and the constant-roll scenarios
were examined separately. In the context of this scalar-tensor theory, inflation can be viable for both scenarios since the observational indices take acceptable values according to the most recent Planck data. Furthermore, the involvement of the
Chern-Simons term has no effect on the background equations, in contrast to the scalar function which couples with the Ricci scalar and participates in the equations of motion. However, the Chern-Simons term ensures the chirality of stochastic gravitational waves. A blue-tilted tensor spectral index of primordial curvature perturbations can be manifested since, tensor modes are strongly affected by the Chern-Simons term. Lastly, the Swampland criteria and the Lyth-bound were examined in order to distinguish the effective field theories towards the path of a consistent M-theory.
\end{abstract}


\maketitle
\section{Introduction}
The inflationary era of the primordial Universe constitutes a quite mysterious era since, the exact mechanism that led the Universe into an accelerated expansion is still unknown. According to the pioneer scenario, which was first proposed by Alan Guth \cite{Guth:1980zm} and formed the background for the development of many alternative modifications, the Universe experienced an abrupt expansion soon after the Big Bang. The inflationary paradigm describes the cosmological evolution of the early Universe and provides a sufficient explanation about many issues, that arise from the well-established Big Bang Cosmological Model. Specifically, inflation provides an explanation about the flatness and the horizon problem of the Universe, the absence of primordial relics and the problem of the initial entropy. 

Until now, the Lagrangian of the inflationary era has not been determined by observational constraints. This fact allows us to interpret that the Primordial era of the Universe can be characterized as classical, considering the Einstein's theoretical framework with the involvement of quantum fluctuations around a quasi de Sitter background Ref. \cite{Cheung:2007st}. This indicates that the inflationary epoch succeeds the quantum era, that was dictated by a quantum gravitational theory and preceded by the classical era. Based on the aforementioned assumption it is plausible to consider that the imprint of the quantum era may have remained during the inflationary era with the inclusion of higher curvature terms that originate from higher dimensional theories of gravity such as Superstring theory. One of these string-inspired corrections is the Chern-Simons parity violating term, which was extensively examined in the literature, for further details see Refs. \cite{Nojiri:2020pqr,Nojiri:2019nar,Odintsov:2019evb,Odintsov:2019mlf,Alexander:2009tp,Qiao:2019hkz,Nishizawa:2018srh,Wagle:2018tyk,Yagi:2012vf,Yagi:2012ya,Molina:2010fb,Izaurieta:2009hz,Smith:2007jm,Konno:2009kg,Sopuerta:2009iy,Matschull:1999he,Haghani:2017yjk,Fronimos:2021czc,Venikoudis:2021oee}. An interesting feature is that the Chern-Simons model violates parity, or in other words the tensor modes behave differently depending on their chirality. The fact that circular polarizations behave differently, leaves an imprint on the tensor spectral index and the tensor-to-scalar ratio which in turn implies that models which were previously incompatible with observations can now become viable, in principle. 

In addition, the upcoming years seem to be quite promising for modern cosmologists since, many earth and space based interferometric gravitational wave experiments will attempt to detect stochastic signals from gravitational wave background
opening new frontier in theoretical Cosmology. The General Theory of Relativity with the inclusion of a scalar field seems to produce an energy spectrum of primordial gravitational waves that is quite weak in the large frequency regime, which concerns modes that behave sub-horizon in the early era. One can speculate that if an enhanced signal were to be detected the following years, then it might originate from a modified gravitational theory or a scalar-tensor theory. For instance a theory involving a gravitational Chern-Simons term may produce a detectable signal and even a blue-tilted tensor spectral index of primordial perturbations.
Recently, experiments on  pulsar timing arrays from NANOGrav have shown that a blue-shifted tensor spectral index can be manifested, for further details see the Refs. \cite{Vagnozzi:2020gtf,Kuroyanagi:2020sfw}. It has also been demonstrated, that the chirality of the primordial gravitational waves is ensured in the context of Einstein-Chern-Simons theories of gravity gravity, according to the Ref. \cite{Odintsov:2022hxu}.

Based on the aforementioned reasoning, the major goal of this work is to examine the inflationary phenomenology of non-minimally coupled Einstein-Chern-Simons gravity and investigate under which conditions, blue stochastic gravitational waves can be generated \cite{Camerini:2008mj,Kuroyanagi:2014nba}. Concerning the involvement of the Chern-Simons corrective term, it does not affect the gravitational equations of motion, however it affects strongly the tensor perturbations of the theory and produces two circular polarization states of gravitational waves. As a result, the overall phenomenology is quite close to the non-minimal canonical scalar field. 

Lastly, we shall examine briefly the Swampland criteria, that where introduced in Refs. \cite{Vafa:2005ui,Ooguri:2006in} and developed further in Refs. \cite{Palti:2020qlc,Mizuno:2019bxy,Brandenberger:2020oav,Blumenhagen:2019vgj,Wang:2019eym,Palti:2019pca,Cai:2018ebs,Akrami:2018ylq,Mizuno:2019pcm,Aragam:2019khr,Brahma:2019mdd,Mukhopadhyay:2019cai,Yi:2018dhl,Gashti:2022hey,Brahma:2019kch,Haque:2019prw,Heckman:2019dsj,Acharya:2018deu,Elizalde:2018dvw,Cheong:2018udx,Heckman:2018mxl,Kinney:2018nny,Garg:2018reu,Lin:2018rnx,Park:2018fuj,Olguin-Tejo:2018pfq,Fukuda:2018haz,Wang:2018kly,Ooguri:2018wrx,Matsui:2018xwa,Obied:2018sgi,Agrawal:2018own,Murayama:2018lie,Marsh:2018kub,Storm:2020gtv,Trivedi:2020wxf,Sharma:2020wba,Odintsov:2020zkl,Mohammadi:2020twg,Trivedi:2020xlh,Oikonomou:2021zfl,Benetti:2019smr} for particular models, in order to investigate whether the model at hand can safely be considered as an effective model. The presence of the non-minimal scalar coupling function alters the dynamical evolution of the scalar field so it is interesting to examine how the Swampland criteria are affected by such assumption. Of course the Chern-Simons scalar coupling function does not affect the viability of the aforementioned criteria given that only tensor modes are influenced by such contribution.  

The structure of this paper is the following; In Sec. II we develop the theoretical framework of the inflationary phenomenology of a non-minimally coupled Einstein-Chern-Simons gravitational theory. In Sec. III we present under which circumstances an inflationary model can be characterized as viable based on the compatibility with the most recent Planck data. In addition, the Swampland Criteria are demonstrated. Finally, Section IV involves two models that are compatible with observations under the slow-roll assumption while, in Sec.V the constant-roll condition is showcased. Last but not least, the concluding remarks are showcased.

\section{Theoretical Framework of non-minimally coupled Chern-Simons Gravity}
The model we propose is described in the context of classical Einstein's gravity in the presence of scalar field, which represents low-effective energy corrections and the string-inspired Chern-Simons parity violating term. The key feature of the theory is the inclusion of a scalar coupling function non-minimally coupled with the Ricci scalar. Specifically, the gravitational action reads \cite{Hwang:2005hb},
\begin{widetext}
\begin{equation}
\centering
\label{action}
\mathcal{S}=\int{d^4x\sqrt{-g}\left(\frac{h(\phi)R}{2\kappa^2}-\frac{1}{2}g^{\mu\nu}\nabla_\mu\phi\nabla_\nu\phi 
-V(\phi)+ 
\frac{1}{8}\nu(\phi)R\tilde R\right)}\, ,
\end{equation}
\end{widetext}
with $h(\phi)$ being a dimensionless scalar function, which couples to the Ricci scalar, g stands for the determinant of the metric tensor and $\frac{1}{2}g^{\mu\nu}\nabla_\mu\phi\nabla_\nu\phi$, $V(\phi)$ represent the scalar kinetic term and the inflaton's potential respectively. For the sake of simplicity, a canonical scalar field shall be assumed however, one can draw conclusions about the phantom case by simply inverting the sign of the kinetic term in the following computations. Furthermore, the scalar coupling function $h(\phi)$ is assumed to be strictly positive for the sake of consistency. The curvature invariant is given by the following expression $R\Tilde{R}=\epsilon^{\mu \nu \sigma \rho}R^{\ \ \alpha \beta }_{\mu \nu}R_{\sigma \rho \alpha \beta}$ where, $\epsilon^{\mu \nu \sigma \rho}$ is the Levi-Civita totally antisymmetric tensor in 4D. Lastly, $\nu(\phi)$ represents the Chern-Simons scalar coupling function. In addition, it is considered that the geometric background is a flat Friedmann–Lemaître–Robertson–Walker background, with the line element of the form, 

\begin{equation}
\centering
\label{metric}
ds^2=-dt^2+a^2(t)\delta_{ij}dx^idx^j\, ,
\end{equation}
where $a(t)$ stands for scale factor of the Universe and the metric tensor has the following form $g_{\mu\nu}=diag(-1, a^2(t), a^2(t), a^2(t))$. In consequence, it is reasonable to assume that the scalar field is also homogeneous in order to obtain time-dependent field equations. As a result, the kinetic term of the scalar field is further simplified as now $\frac{1}{2}g^{\mu\nu}\nabla_\mu\phi\nabla_\nu\phi=-\frac{1}{2}\dot\phi^2$. Let us now see how the above assumption simplifies the inflationary phenomenology.

In order to produce the field equations of the theory, we apply the variational principle into the aforementioned gravitational action (\ref{action}) with respect to the metric tensor and the canonical scalar field. The former generates the gravitational field equations, which hold information about the evolution of the primordial universe while, the latter describes the dynamical evolution of the scalar field. Obviously, both cases are interconnected and this can easily be inferred from the following system of equations. Firstly, the gravitational field equations for the non-minimal Chern-Simons model now read \cite{Hwang:2005hb},

\begin{equation}
\centering
\label{fieldeq}
\frac{h(\phi)}{\kappa^2}G_{\mu\nu}=\nabla_\mu\phi\nabla_\nu\phi-\bigg[\frac{1}{2}g^{\alpha\beta}\nabla_\alpha\phi\nabla_\beta\phi+V\bigg]g_{\mu\nu}+T^{(CS)}_{\mu\nu}\, ,
\end{equation}
where $T^{(CS)}_{\mu\nu}=\frac{1}{4\sqrt{-g}}\frac{\delta(\sqrt{-g}\nu(\phi)R\tilde R)}{\delta g^{\mu\nu}}$ is the stress tensor for the Chern-Simons model which reads,

\begin{equation}
\centering
\label{TCS}
T_{\mu\nu}^{(CS)}=\epsilon_{\mu}^{\,\,\,\alpha\beta\gamma}(\nu_{,\gamma;\delta}R^\delta_{\,\,\nu\alpha\beta}-2\nu_{,\gamma}R_{\nu\alpha;\beta})\, ,
\end{equation}
while the scalar field satisfies the Klein-Gordon equation,

\begin{equation}
\centering
\label{KGeq}
\Box\phi-\frac{dV}{d\phi}+\frac{R}{2\kappa^2}\frac{dh}{d\phi}=0\, .
\end{equation}
Note that Eq. (\ref{TCS}) implies that the energy stress tensor vanishes when the scalar field does not evolve dynamically. Furthermore, since it is homogeneous, only spatial components survive. Also, in this approach, it can easily be inferred that due to the homogeneous and isotropic background, the Chern-Simons term does not participate in the above relations, therefore in the end, the generated system of equations is,
\begin{equation}
\centering
\label{motion1}
\frac{3hH^2}{\kappa^2}=\frac{1}{2}\dot\phi^2+V-\frac{3H\dot h}{\kappa^2}\, ,
\end{equation}

\begin{equation}
\centering
\label{motion2}
-\frac{2h\dot H}{\kappa^2}=\dot\phi^2+\frac{\ddot h-H\dot h}{\kappa^2}\, ,
\end{equation}

\begin{equation}
\centering
\label{motion3}
\ddot\phi+3H\dot\phi+V'-\frac{R}{2\kappa^2}h'=0\, ,
\end{equation}
where the prime represents differentiation with respect to the scalar field. The above set of equations is quite intricate in order to be solved analytically specifically, due to the involvement of the scalar coupling function $h(\phi)$ and its derivatives. It is also worth mentioning that even though the Chern-Simons term is absent from the equations of motion, its influence on the primordial tensor perturbations is strong as we shall showcase subsequently. At this point of our analysis, we apply the slow-roll approximations during the inflationary era, in order to further simplify the set of equations (\ref{motion1}-\ref{motion3}). More specifically, in order to obtain a viable inflationary era with a sufficient amount of expansion so that the apparent flatness and horizon issues are resolved, a slow-rolling scalar field shall be assumed during the first horizon crossing. Hence, the kinetic term of the scalar field is assumed to be inferior compared to the scalar potential, or in other words $\frac{1}{2}\dot\phi^2\ll V$. In consequence, by differentiating the above relation, it can easily be inferred that $\ddot\phi\ll V'$ or according to the continuity equation of the scalar field (\ref{motion3}), $\ddot\phi\ll H\dot\phi$. In consequence, by applying the slow-roll assumption to every scalar function of the model,  the following approximations shall be considered,
\begin{align}
\centering
\label{slowroll}
\dot H&\ll H^2\, ,&\frac{1}{2}\dot\phi^2&\ll V\, ,& \ddot\phi&\ll H\dot\phi\, ,&\ddot h&\ll H\dot h\, .
\end{align}
As a result, the background equations can be approximated as,
\begin{equation}
\centering
\label{motion4}
H^2\simeq\frac{\kappa^2 V}{3h}\, ,
\end{equation}

\begin{equation}
\centering
\label{motion5}
\dot H\simeq-\frac{\kappa^2\dot\phi^2}{2h}+\frac{H\dot h}{2h}\, ,
\end{equation}

\begin{equation}
\centering
\label{motion6}
3H\dot\phi+V'-\frac{R}{2\kappa^2}h'\simeq0\, ,
\end{equation}
where the Ricci scalar can be written as $R=6\dot H+12H^2$ and based on the slow-roll approximation its reduced form is $R\simeq 12H^2$. It is interesting to observe that the non-minimal scalar coupling function $h(\phi)$, apart from appearing in the Friedmann equation, it affects the Raychaudhuri equation through the term $\dot h$ assuming that $h$ evolves dynamically until the scalar field reaches its vacuum expectation value. One can now easily solve the continuity equation (\ref{motion6}) with respect to the time derivative of the scalar field and get the following result,
\begin{equation}
\centering
\label{fdot}
\dot\phi\simeq-\frac{V'-\frac{6H^2}{\kappa^2}h'}{3H}\, ,
\end{equation}
which is a necessary quantity for the overall phenomenological description. In this approach, it can easily be inferred that the non-minimal coupling function can become dominant compared to the scalar potential, under proper circumstances, therefore depending on the coupling functions, the phenomenology can be quite different. Now, before we proceed with the derivation of the spectral indices, let us see how the Chern-Simons scalar coupling function affects tensor perturbations.

We commence by assuming that the metric in Eq.(\ref{metric}) is affected by the presence of tensor perturbations as,

\begin{equation}
\centering
\label{perturbedmetric}
ds^2=-dt^2+a^2(t)(\delta_{ij}+h_{ij}(t,\bm x))dx^idx^j\, ,
\end{equation}
where $h_{ij}(t,\bm x)$ describes tensor perturbations. For consistency, it should be stated that in order to properly describe tensor perturbations, the aforementioned function describes only traceless transverse modes, that is $\delta^{ij}h_{ij}=0=\partial_i h^{ij}$. Hence, it can easily be proven that tensor modes must satisfy the following relation \cite{Hwang:2005hb}, 
\begin{widetext}
\begin{equation}
\centering
\label{tensorperturbations}
\frac{1}{ha^3}\frac{d}{dt}\bigg(a^3h \dot h_{ij}\bigg)-\frac{\nabla^2}{a^2}h_{ij}
 -\frac{2k^2}{ha}\epsilon_{(i}^{\ \ kl}\bigg[(\ddot\nu-H\dot\nu)\dot h_{j)k}-\dot\nu \Box h_{j)k}\bigg]_{,l}=0\, ,
\end{equation}
\end{widetext}
where $-\Box h_{ij}=\ddot{h}_{ij}+3H\dot{h}_{ij}-\frac{\nabla^2}{\alpha^2}h_{ij}$. In consequence, by performing a Fourier transformation on the tensor perturbation of the form,
\begin{equation}
\centering
\label{ansatz}
h_{ij}(t,\bm x)=\sqrt{Vol}\int \frac{d^3\bm k}{(2\pi)^3}\sum_{l}e_{\alpha\beta}^{(l)}(\bm k)h_{l\bm{k}}(t)e^{i\bm{k}\cdot\bm{x}},
\end{equation}
then one can diagonalize the equation of tensor perturbations in momentum space as shown below,
\begin{equation}
    \centering
    \label{modeeq}
    \frac{1}{a^3Q_{t}}\frac{d}{dt}\left(a^3Q_{t}\dot h_{l \bm{k}}\right)+c_\mathcal{T}^2\frac{k^2}{a^2}h_{l \bm{k}}\, ,
\end{equation}
where the Chern-Simons auxiliary function $Q_t$ is defined as,
\begin{equation}
\centering
\label{Qt}
    Q_t=\frac{h}{\kappa^2}+2\lambda_l\dot\nu \frac{k}{a}\, ,
\end{equation}
and involves the scalar coupling function $h(\phi)$. Here, it should be stated that $e_{ij}^{(l)}$ is the circular polarization state of tensor modes and $\lambda_l$ is an auxiliary index that takes the values $\lambda_l=\pm1$ for right and left handed polarization states respectively. Since we are interested in the first horizon crossing and due to the fact that the sound wave velocity is equal to unity, one can effectively perform the replacement $\frac{k}{a}=H$. In addition, the propagation velocity of tensor perturbations is not affected by the inclusion of the Chern-Simons term therefore $c_\mathcal{T}=1$. As it is obvious, the involvement of the Chern-Simons term affects the tensor perturbations of the theory since, it differentiates between different types of circular polarizations. This can easily be inferred from the fact that based on their circular polarization, modes must satisfy a different differential equation. As expected, this has an impact on tensor modes and this can be quantified in the tensor spectral index of the model. Let us see how one can obtain information about the spectral indices.

During the inflationary era, the cosmological evolution is governed by the slow-roll parameters where, according to Ref. \cite{Hwang:2005hb} their expressions are given by the following relations,
\begin{align}
\centering
\label{indices}
\epsilon_1&=-\frac{\dot H}{H^2}\, ,& \epsilon_2&=\frac{\ddot\phi}{H\dot\phi}\, ,&
\epsilon_3&=\frac{\dot h}{2Hh}\, ,&
\epsilon_4&=\frac{\dot E}{2HE}\, ,\nonumber\\
\epsilon_5&=\frac{1}{2}\sum_l\frac{\dot Q_t}{2HQ_t}\, ,
\end{align}
where the index $\epsilon_4$ is determined by the auxiliary function E of the form,
\begin{equation}
\centering
E=\frac{h}{\kappa^2}\left(1+\frac{3\dot h^2}{2h\kappa^2\dot\phi^2}\right)\, .
\end{equation}
It becomes clear that the impact of the Chern-Simons term on the inflationary phenomenology lies exclusively on the slow-roll index $\epsilon_5$ and the summation is over the left and right handed polarization states of primordial gravitational waves. In addition, the rest slow-roll indices are used in order to take into consideration the impact of the non-minimal scalar coupling function and the scalar potential. In the limit of a vanishing scalar function $h(\phi)$, indices $\epsilon_3$ and $\epsilon_4$ vanish identically. Last but not least, the first slow-roll index shall be used in order to ascertain the conditions under which inflation terminates and afterwards, conclusions shall be drawn about the viability of a given model. Furthermore, for simplicity it is more convenient to express the indices $\epsilon_4-\epsilon_5$ using the indices $\epsilon_1-\epsilon_3$ and the auxiliary dimensionless functions x and y as demonstrated below,
\begin{equation}
\centering
\label{index4}
\epsilon_4\simeq\frac{\epsilon_3}{x^2+\epsilon_3^2}\bigg(x^2+\frac{h''\dot\phi}{Hh'}\epsilon_3\bigg)\, ,
\end{equation}

\begin{equation}
\centering
\label{index5}
\epsilon_5\simeq\frac{1}{2}\sum_l\bigg(\frac{\epsilon_3+\frac{y}{2}(\frac{\nu''\dot\phi}{H\nu'}+\epsilon_2-\epsilon_1)}{1+y}\bigg)\, ,
\end{equation}
where $x=\frac{\kappa\dot\phi}{\sqrt{6h}H}$ and $y=\frac{2\lambda_l\kappa^2\dot\nu H}{h}$ are auxiliary dimensionless parameters. Lastly, we showcase the generalized expressions of the slow-roll indices for arbitrary scalar functions $h(\phi)$, $V(\phi)$ and $\nu(\phi)$,

\begin{equation}
\label{epsilon1general}
\centering
\epsilon_1 \simeq \frac{\left(h V'-2 V h'\right) \left(h V'-V h'\right)}{2 \kappa ^2 h V^2}\, ,
\end{equation}

\begin{equation}
\label{epsilon2general}
\centering
\begin{split}
\epsilon_2 \simeq \frac{h V \left((4 h-3) h' V'-2 h V''\right)}{2 \kappa ^2 h V^2} \\
+\frac{2 V^2 \left(h'^2+2 h^2 h''\right)+h^2 V'^2}{2 \kappa ^2 h V^2} \, ,
\end{split}
\end{equation}

\begin{equation}
\label{epsilon3general}
\centering
\epsilon_3 \simeq \frac{h' \left(\frac{2 h'}{h}-\frac{V'}{V}\right)}{2 \kappa ^2} \, .
\end{equation}
According to the above equations, it can easily be inferred that the 
scalar coupling function $h(\phi)$ has a major role on the overall phenomenology since, it participates effectively in all of the slow-roll parameters. In addition, letting $h\to 1$ generates the known expressions for the minimal model.

\section{Swampland Criteria And Energy Spectrum of Gravitational Waves}
For an inflationary model to be characterized as viable, it must be consistent with the latest observational constraints. According to the latest Planck data, which are available in Ref. \cite{Planck:2018jri}, the observed indices, namely the scalar spectral index of primordial perturbations and the tensor-to-scalar-ratio have the following numerical values $n_\mathcal{S}=0.9649\pm 0.0042$ with 68$\%$ confidence level and the tensor-to-scalar ratio must satisfy the condition $r<0.064$ with 95$\%$ confidence level. To this day, the exact value of the tensor spectral index of primordial perturbations or at least, the numerical spectrum in which it resides is still unknown due to the fact that B-modes \cite{Kamionkowski:2015yta} have yet to be detected. This fact is quite intriguing and at the same a challenging topic that may be solved in the upcoming years from third generation interferometers, that will map the Universe in mHz frequencies. An interesting trait of the participation of the Chern-Simons term in the gravitational action is that it can produce a positive tensor spectral index as we shall demonstrate subsequently. The observed indices can be written in terms of the previously defined slow-roll indices and their expressions are given as follows \cite{Hwang:2005hb},
\begin{align}
\centering
\label{observables}
n_\mathcal{S}&=1-2\frac{(2\epsilon_1+\epsilon_2-\epsilon_3+\epsilon_4)}{1-\epsilon_1}\, ,&n_\mathcal{T}&=-2\frac{(\epsilon_1+\epsilon_5)}{1-\epsilon_1}\, ,&\nonumber\\r&=8|\epsilon_1+\epsilon_3|\sum_l\bigg|\frac{1}{1+y}\bigg|\, ,
\end{align}
where as it obvious, they are strongly dependent on the slow-roll indices.
Note that in this form, $\epsilon_1+\epsilon_3\simeq\frac{1}{2h}\left(\frac{\kappa\dot\phi}{H}\right)^2$ according to Eq. (\ref{motion5})  while the auxiliary function y has been presented in the previous section. In this case, it can easily be inferred that as long as the condition $\epsilon_5<-\epsilon_1$ is satisfied, then the tensor spectral index obtains a positive value. In this approach, the above inequality is model dependent however, in principle it can be satisfied while the scalar spectral index and the tensor-to-scalar ratio obtain a compatible with observations value. This will become apparent in the following section. Note also that the numerical value of a positive tensor spectral index cannot be arbitrarily large in order to be attributed to a stochastic gravitational wave background and the following upper bound $n_\mathcal{T}<0.5$ must be respected \cite{Giare:2020vss}. 

Let us now describe the necessary steps that will be used in order to ascertain whether a particular model can be regarded as viable. We commence with  the e-folding number which indicates the duration of the inflationary epoch and under the assumption that a scalar field dominates primordially, it has the following form,
\begin{equation}
\centering
\label{efolds}
N=\int_{\phi_k}^{\phi_{end}}{d\phi\frac{H}{\dot \phi}}\, ,
\end{equation}
with $\phi_k$ being the initial value of the scalar field during the first horizon crossing and $\phi_{end}$ denoting the inflaton's final value when the inflationary epoch ceases. It is expected that the numerical spectrum of the e-folding number ranges in the area $50-60$ approximately in order to resolve the apparent flatness and horizon problems. One can find the value $\phi_{end}$ for a model of interest by setting the first slow-roll index in Eq. (\ref{epsilon1general}) equal to unity and then substitute it into Eq. (\ref{efolds}) in order to extract the initial value $\phi_k$. However, since the first index $\epsilon_1$ has a complicated form and at the same time, the first time derivative of the scalar field, which participates in the denominator of the e-folding equation is also perplexed one can expect that exporting the value of $\phi_k$ is not an easy task, especially for complicated scalar potentials. This is the reason why hereafter, potentials with power-law form in the models shall be considered.

At this point, it is interesting to examine the conditions that differentiate between UV complete and effective models. In order to extract information about the effectiveness of the model, the Swampland criteria shall be investigated \cite{Vafa:2005ui,Ooguri:2006in}. When the aforementioned criteria are satisfied, this is indicative of the effectiveness of a model. The Swampland criteria have been thoroughly studied in the literature \cite{Palti:2020qlc,Mizuno:2019bxy,Brandenberger:2020oav,Blumenhagen:2019vgj,Wang:2019eym,Palti:2019pca,Cai:2018ebs,Akrami:2018ylq,Mizuno:2019pcm,Aragam:2019khr,Brahma:2019mdd,Mukhopadhyay:2019cai,Yi:2018dhl,Gashti:2022hey,Brahma:2019kch,Haque:2019prw,Heckman:2019dsj,Acharya:2018deu,Elizalde:2018dvw,Cheong:2018udx,Heckman:2018mxl,Kinney:2018nny,Garg:2018reu,Lin:2018rnx,Park:2018fuj,Olguin-Tejo:2018pfq,Fukuda:2018haz,Wang:2018kly,Ooguri:2018wrx,Matsui:2018xwa,Obied:2018sgi,Agrawal:2018own,Murayama:2018lie,Marsh:2018kub,Storm:2020gtv,Trivedi:2020wxf,Sharma:2020wba,Odintsov:2020zkl,Mohammadi:2020twg,Trivedi:2020xlh,Oikonomou:2021zfl,Benetti:2019smr} and interesting phenomenological implications have been showcased in various modified theories of gravity, see for instance  \cite{Benetti:2019smr} for an interesting application on $f(R)$ gravity by making use of Noether's theorem. Concerning the Swampland Criteria, the theory must satisfy the following relations,
\begin{itemize}
  \item The first criterion is the Swampland distance conjecture. According to this, the effective theory must have a field range of the following form,
  \begin{equation}
  \centering
  \label{criterion1}
  |\kappa\Delta\phi|<\mathcal{O}(1)\, ,
  \end{equation}
  and is independent of the sign.
  \item The second criterion is the de Sitter conjecture. It speculates that the derivative of the scalar potential has a lower bound,
  \begin{equation}
  \centering
  \label{criterion2}
  \frac{|V'(\phi_k)|}{\kappa V(\phi_k)}>\mathcal{O}(1)\, ,
  \end{equation}
 or alternatively, the scalar potential can satisfy the following relation, 
  \begin{equation}
  \centering
  \label{criterion3}
  -\frac{V''(\phi_k)}{\kappa^2V(\phi_k)}>\mathcal{O}(1)\, ,
  \end{equation}
\end{itemize}
where the prime denotes differentiation with respect to the scalar field $\phi$. It should be stated that in order to ascertain whether the model at hand is effective, it is not required for the criteria to be satisfied simultaneously. In particular, if a single criterion is satisfied then the model can be regarded as effective. On the other hand, there exist instances where the criteria are actually at variance and cannot be satisfied at the same time. Consider for instance the power-law model in Ref. \cite{Oikonomou:2021zfl}, which it was shown that, when the second criterion is respected, its supplementary form cannot be satisfied.  

Before we proceed with the numerical analysis of a few models, let us first investigate the impact that a blue spectrum has on the energy spectrum of primordial gravitational waves. Previously, it has shown that in momentum space, tensor modes, depending on their circular polarization, must satisfy the following differential equation \cite{Odintsov:2022cbm,Hwang:2005hb}, 

\begin{equation}
\label{tensormodes}
\centering
\ddot h(k)+(3+\alpha_M)H \dot h(k)+\frac{k^2}{\alpha^2}h(k)=0,
\end{equation}
where $\alpha_M=\frac{\dot Q_t}{HQ_t}$ denotes the running Planck mass \cite{Linder:2021pek,Mitra:2020vzq,DAgostino:2019hvh,Nunes:2018zot} for a specific circular polarization. Based on the above expression, one can infer that $\alpha_M$ carries the impact of the respective modified gravity with the inclusion of other string-corrective terms on the evolution of primordial gravitational waves. In this theory, both the non-minimal coupling function and the Chern-Simons term affect tensor modes. Hence, it can be shown that the running Planck mass must have the following expression,
\begin{equation}
\label{a_M}
\centering
\alpha_M=\frac{\dot h}{Hh}+\frac{\frac{2\lambda_l\kappa^2\dot\nu}{h}\frac{k}{a}}{1+\frac{2\lambda_l\kappa^2\dot\nu}{h}\frac{k}{a}}\bigg[\frac{\ddot\nu}{H\dot\nu}-\frac{\dot h}{Hh}-1\bigg]\, .
\end{equation}
The effects of modified theories of gravity are actually significant in the energy spectrum of primordial gravitational waves. This can be seen from the computation of the energy spectrum. For GR, it becomes clear that the energy spectrum is given by the following expression,

\begin{equation}
\centering
\label{OmegaGR}
\Omega_{GW}(k)=\frac{1}{\rho_{crit}}\frac{d\bra{0}\rho_{GW}\ket{0}}{d\ln k}=\frac{k^2\Delta_h^2(k)}{12H_0^2}\, ,
\end{equation}
where $H_0$ denotes the current value of the Hubble rate expansion and $\Delta_h^2$ denotes the tensor power spectrum. In particular, according to Ref.\cite{Odintsov:2022cbm} the tensor power spectrum reads,

\begin{align}
\centering
\label{tensorpowerspectrum}
\Delta_h^2(k)&=\Delta_h^{(p)2}(k)\bigg(\frac{\Omega_m}{\Omega_\Lambda}\bigg)^2\bigg(\frac{g_*(T_{in})}{g_{*0}}\bigg)\bigg(\frac{g_{*s0}}{g_{*s}(T_{in})}\bigg)^{\frac{4}{3}}\nonumber\\&\bigg(\overline{\frac{3j_1(k\tau_0)}{k\tau_0}}\bigg)^2T_1^2(x_{eq})T_2^2(x_R)\, ,
\end{align}
where $x_i=k/k_i$ denotes the ratio between a wavenumber $k$ with a specific wavenumber $k_i$, in the above expressions subscripts "eq" and "R" denote the matter-radiation equivalence and the reheating era respectively, $T$ is the transfer function, $\Omega_m$ and $\Omega_\Lambda$ are the density parameters for matter and dark energy, $g$ denotes the relativistic degrees of freedom which, depending on the temperature, or in other words the cosmological era, it alters and finally $\Delta_h^{(p)2}$ denotes the primordial tensor power spectrum. In this approach, it can easily be inferred that,

\begin{equation}
\centering
\label{primordialspectrum}
\Delta_h^{(p)2}(k)=r\mathcal{P}_\zeta\bigg(\frac{k}{k_*}\bigg)^{n_\mathcal{T}}\, ,
\end{equation}
where $k_*$ is the CMB pivot scale and in principle, the tensor spectral index in the exponent is depending on the frequency, or in other words $n_\mathcal{T}=n_{\mathcal{T}}(k)$. For the sake of generality, one can expand the tensor spectral index around the pivot scale in a power series as \cite{Zarei:2014bta},

\begin{equation}
\centering
\label{nTexpand}
n_\mathcal{T}(k)=n_\mathcal{T}(k_*)+\sum_{n=1}^\infty\frac{d^n\ln n_{\mathcal{T}}}{d\ln k^n}\bigg|_{k_*}\frac{\ln^n\frac{k}{k_*}}{(n+1)!}\, ,
\end{equation}
however due to the fact that higher powers are subleading, one can simply work on the leading order $a_\mathcal{T}(k_*)=\frac{dn_\mathcal{T}}{d\ln k}$. In the end, one can show that the energy spectrum for GR must have the following form,

\begin{align}
\centering
\label{GWGRspectrum}
\Omega_{GW}&=\frac{k^2}{12H_0^2}r\mathcal{P}_{\zeta}\bigg(\frac{k}{k_*}\bigg)^{n_{\mathcal{T}}(k_*)+\frac{a_\mathcal{T}(k_*)}{2}\ln\frac{k}{k_*}}\bigg(\frac{\Omega_m}{\Omega_\Lambda}\bigg)^2\bigg(\frac{g_*(T_{in})}{g_{*0}}\bigg)\nonumber\\&\bigg(\frac{g_{*s0}}{g_{*s}(T_{in})}\bigg)^{\frac{4}{3}}\bigg(\overline{\frac{3j_1(k\tau_0)}{k\tau_0}}\bigg)^2T_1^2(x_{eq})T_2^2(x_R)\, .
\end{align}
Due to the fact that the second transfer function carries information about the reheating temperature, studying the energy spectrum of gravitational waves is quite promising since the high frequency regime can now shine light towards the currently unknown cosmological eras right after inflation since these modes are the first to re-enter the horizon. Now, in order to examine the impact that modified theories of gravity have, one needs to solve Eq. (\ref {tensormodes}). In order to do so, it is convenient to make use of the WKB approximation \cite{Nishizawa:2011eq}, therefore the tensor mode can be written with respect to the GR result as \cite{Odintsov:2022cbm},

\begin{equation}
\centering
\label{hMG}
h_{M}=\e^{-\mathcal{D}}h_{GR}\, ,
\end{equation}
where $\mathcal{D}=\frac{1}{2}\int_0^zdz'\frac{a_M}{1+z'}$ and $h_{ij}=e_{ij}h_M$. This in turn implies that the exponent $\mathcal{D}$ that contains information about the modified gravity appears in the energy spectrum of gravitational waves through the tensor power spectrum. Since it appears in a square form, one can show that \cite{Odintsov:2022cbm},

\begin{widetext}
\begin{equation}
\centering
\label{OmegaMG}
\Omega_{GW}=\e^{-2\mathcal{D}}\frac{k^2}{12H_0^2}r\mathcal{P}_{\zeta}\bigg(\frac{k}{k_*}\bigg)^{n_{\mathcal{T}}(k_*)+\frac{a_\mathcal{T}(k_*)}{2}\ln\frac{k}{k_*}}\bigg(\frac{\Omega_m}{\Omega_\Lambda}\bigg)^2\bigg(\frac{g_*(T_{in})}{g_{*0}}\bigg)\bigg(\frac{g_{*s0}}{g_{*s}(T_{in})}\bigg)^{\frac{4}{3}}\bigg(\overline{\frac{3j_1(k\tau_0)}{k\tau_0}}\bigg)^2T_1^2(x_{eq})T_2^2(x_R)\, .
\end{equation}
\end{widetext}
Hence, compared to the GR result, the energy spectrum of gravitational waves can be amplified for $\mathcal{D}<0$ and also, in the high frequency regime, a positive tensor spectral index results in an amplification as well. The effect of a parity odd term in the gravitational action (\ref{action}) on the energy spectrum of gravitational waves is also quite different from GR due to the fact that chiral gravitational waves are now expected. More specifically, due to the fact that each circular polarization satisfies a unique differential equation, the running Planck mass in principle differs, recall Eq. (\ref{a_M}) for $\lambda_l=\pm 1$. Depending on the coupling functions chosen, the distinction between left and right handed polarization states may be significant to the point where for a given frequency, two signals are expected in the energy spectrum, see for instance  \cite{Odintsov:2021kup}. In the non-minimal Chern-Simons model, the enhancement that is predicted for the energy spectrum applies only to high frequencies in which the scalar field evolves dynamically with respect to cosmic time. In other words, low-frequencies should generate the same result as GR regardless of the polarization since the scalar field in this case has reached its vacuum expectation value. Hence, on the late-time, no distinction between polarization is expected. In the following, we shall examine a few power-law models in order to ascertain whether a viable inflationary era that enhances the high frequency regime of primordial gravitational waves is possible.

\section{Confronting Models of interest with observations}

\subsection{Trigonometric Chern-Simons Coupling}
In our first proposed model the inflaton's scalar potential is given by the following linear form,
\begin{equation}
\centering
\label{V1}
V(\phi)=V_1(\kappa\phi)\, ,
\end{equation}
where, the amplitude of the potential has dimensions $[V_1]=$eV$^4$ for the sake of completeness. We consider that the scalar function, which couples with the Ricci scalar is,
\begin{equation}
\centering
\label{h1}
h(\phi)=q(\kappa \phi)^{-1}\, ,
\end{equation}
where the free parameter q can be considered as dimensionless. This model is chosen so that the second criterion for the Swampland conjecture can be determined only by the first form while, provided that $\phi_k<M_P$, $h(\phi_k)$ dominates over the potential if the amplitudes are proper. Furthermore, when the vacuum expectation value for the scalar field is reached, the potential can behave as an effective cosmological constant in the late-time. Lastly, the Chern-Simons scalar coupling function is defined as follows,
\begin{equation}
\centering
\label{v1}
\nu(\phi)=\sin\bigg(p(\kappa \phi)\bigg)\, ,
\end{equation}
where $p$ is a dimensionless parameter. We proceed with the cosmological evolution during the inflationary era, which is governed by the slow-roll parameters as follows,

\begin{equation}
\centering
\label{e1model1}
\epsilon_1\simeq \frac{3 q}{\kappa ^3 \phi ^3}\, ,
\end{equation}

\begin{equation}
\centering
\label{e2model1}
\epsilon_2\simeq  \frac{q (3 \kappa  \phi +2 q)}{\kappa ^4 \phi ^4}\, ,
\end{equation}

\begin{equation}
\centering
\label{e3model1}
\epsilon_3\simeq \frac{\epsilon_1}{2}\, ,
\end{equation}

\begin{equation}
\centering
\label{e4model1}
\epsilon_4\simeq q \left(\frac{6}{\kappa ^3 \phi ^3}-\frac{9}{2 \kappa ^3 \phi ^3+3 q}\right)\, ,
\end{equation}
where we omit to present the index $\epsilon_5$, because it is given by a quite complicated form, which involves trigonometric functions due to the specific definition of the coupling function $\nu(\phi)$. 

One can set the first slow-roll parameter in Eq. (\ref{e1model1}) as equal to unity in order to extract the final value of the inflaton. For the case at hand, this value reads,
\begin{equation}
\centering
\label{ffinal1}
\phi_{end}\simeq \frac{(3q)^{1/3}}{\kappa}\, .
\end{equation}
This is a quite simple form and allows us to easily determine the initial value of the inflaton during the first horizon crossing through the Eq. (\ref{efolds}). In consequence, the resulting expression is,
\begin{equation}
\centering
\label{finitial1}
\phi_k\simeq \frac{(3q+9Nq)^{1/3}}{\kappa}\, .
\end{equation}
As it is obvious from the above relations, the value of the scalar field is strongly depended on the free parameter q, to be specified in the following.

In order to get the numerical results of the model we specify the free parameters. In reduced Planck units, where $\kappa=1$, we set $(V_1, p, q, N)=(100 M_P^4, 100.09, 0.001, 60)$. Based on the aforementioned numerical values the model is compatible with the most recent Planck data since, the scalar spectral index of primordial perturbations is $n_\mathcal{S}=0.966796$ and the tensor-to-scalar ratio takes the value $r=8.1217 \cdot 10^{-9}$. Furthermore a slightly blue-tilted tensor spectral index is manifested since $n_\mathcal{T}=0.000196463$, due to the participation of the Chern-Simons term.
The inflaton's initial value during the first horizon crossing is $\kappa\phi_k=0.815831$
and when the inflation ends reaches the value $\kappa\phi_{end}=0.144225$, therefore the scalar field decreases as time flows by. We also stress that the numerical values of the slow-roll parameters during the first horizon crossing are $\epsilon_1\simeq\epsilon_2=0.00552$, $\epsilon_3\simeq \epsilon_4=0.0027$ and lastly $\epsilon_5=-0.00562309$.

\begin{figure}[h!]
\centering
\includegraphics[width=20pc]{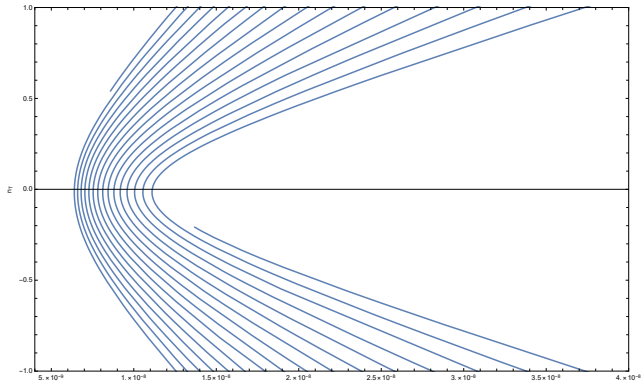}
\caption{Parametric plot of tensor spectral index of primordial perturbations with respect to the tensor-to-scalar-ratio for $N=60$ e-folding. The numerical value of the free parameter q ranges in the vicinity [70, 130].}
\label{parametricA}
\end{figure}

We also highlight that the model can be compatible with the latest Planck data for a wide range of parameters and a red shifted tensor spectral index can be manifested in this context. In particular, the parameter q dictates whether the tensor spectral index will be positive or negative. As an example we consider a small alteration on the parameter q while, the rest parameters remain the same as before. For $q=100$ the tensor spectral index becomes $n_\mathcal{T}=-0.408302$ while, the scalar spectral index remains unchanged and the tensor-to-scalar ratio also satisfies the condition $r<0.064$. In Fig. \ref{parametricA} we demonstrate the behaviour of the 
tensor spectral index as function of the tensor-to-scalar ratio for a wide range of the parameter $q$.

\begin{figure}[h!]
\centering
\includegraphics[width=15pc]{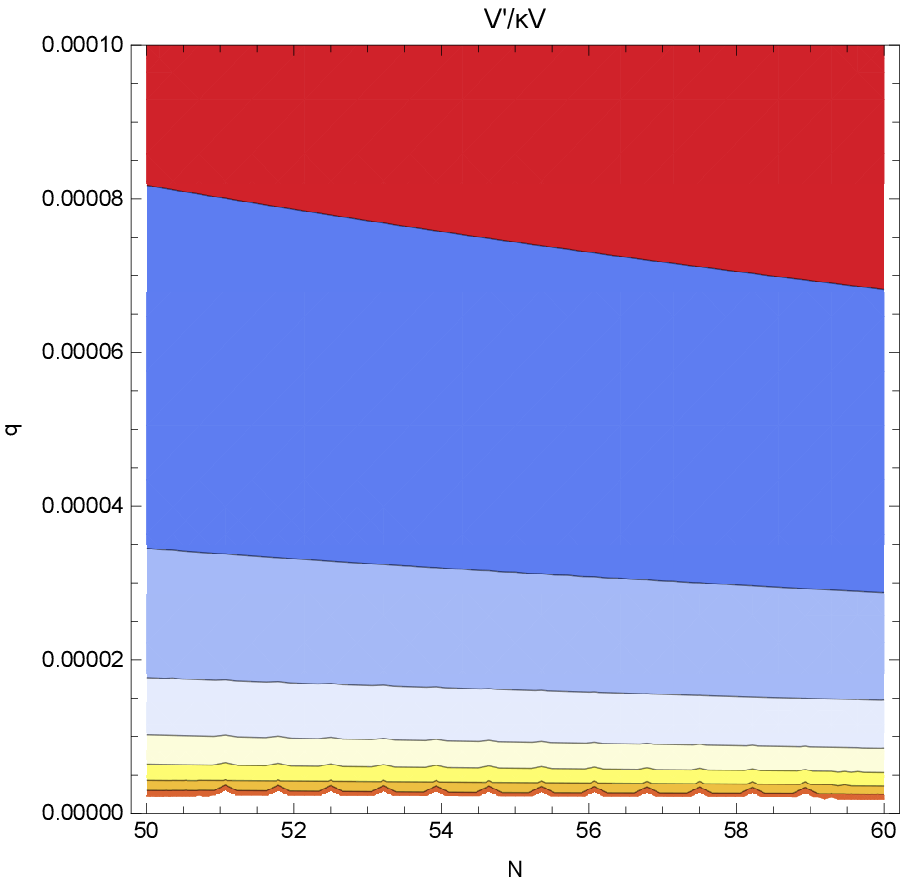}
\includegraphics[width=2pc]{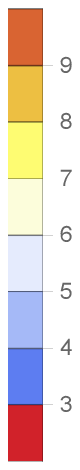}
\caption{Ratio $\frac{V'(\phi_k)}{\kappa V(\phi_k)}$ at the start of inflation depending on the duration of inflation and auxiliary parameter $\lambda$. It can easily be inferred that small values for $\lambda$ manage to increase the numerical value of the ratio.}
\label{Swamp1}
\end{figure}

Concerning the Swampland criteria, one can easily observe that the condition $|\kappa\Delta\phi|<\mathcal{O}(1)$ is valid for the aforementioned set of parameters. In addition, the ratio $\frac{V'(\phi_k)}{\kappa V(\phi_k)}=1.22574$ in the first horizon crossing indicates the satisfaction of the de Sitter conjecture. A more detailed analysis is presented in Fig. \ref{Swamp1} in which it becomes clear that the numerical value of the ratio $\frac{V'}{\kappa V}$ at the start of inflation can in fact become large for small values of parameter $q$. This is a similar approach to the one followed in Ref. \cite{Oikonomou:2021zfl}  however, now the coefficient of the Ricci scalar evolves dynamically with respect to time and freezes when the vacuum expectation value is reached. As it is obvious the alternative form of this criterion has no meaning to be presented since the second derivative of the scalar potential is trivially zero. Therefore, it becomes clear that for the same set of parameters, the Swampland criteria are indeed satisfied.

As a last note we showcase that the slow-roll approximations, which were set in Eq. (\ref{slowroll}) in the first horizon crossing, are indeed satisfied, something which is also hinted from the numerical values of the slow-roll indices. Specifically, 
the approximation $\dot H \ll\ H^2$ holds true since, $H^2\sim \mathcal{O}(10^5)$ and $\dot H\sim \mathcal{O}(10^{3})$. Furthermore, $\frac{1}{2}\dot \phi^2 \sim \mathcal{O}(10^{-1})\ll V\sim \mathcal{O}(10^2)$ and 
$\ddot \phi \sim \mathcal{O}(10^{-1})$, $H\dot \phi \sim \mathcal{O}(10^{2})$ indicating that the rest conditions are also valid. 

Overall, the results are symmetric under the transformation $\phi\to-\phi$, due to the fact that the choice of the non minimal scalar coupling function $h(\phi)$ and the scalar potential results in a Hubble rate symmetric under the aforementioned change even though the coupling functions themselves are antisymmetric.  This in turn implies that a viable inflationary era can also be achieved, consistent with observations using the negative value of the $\phi_{end}$. The numerical results remain the same however, the tensor spectral index takes the value $n_\mathcal{T}=0.142879$.

As a final note, let us briefly discuss the Lyth-bound \cite{Lyth:1996im}. In a sense, it suggests that the field range must be greater than a given threshold,
in order to comply with the observations, or more specifically the fact that primordial gravitational waves have yet to be observed even though they are predicted. In this approach, by considering the form of $\dot\phi$ from Eq. (\ref{fdot}) and the expression for the e-folding number in Eq. (\ref{efolds}), it can easily be inferred that,

\begin{equation}
\centering
\label{Lythbound}
\bigg|h\bigg(\frac{V'}{\kappa V}-2\frac{h'}{\kappa h}\bigg)\bigg|\Delta N<|\kappa\Delta\phi|\, .
\end{equation}
For the model at hand, since $\Delta N=60$ it can be shown that the left hand side is equal to $0.270441$ while, the field range is expected to be $|\kappa\Delta\phi|=0.671606$ therefore, the Lyth-bound in this numerical analysis is respected. It is worth noting that satisfying the Swampland criteria does not spoil the Lyth-bound in this scenario.

\subsection{Quadratic Chern-Simons Coupling}
In this model we consider that the scalar potential is described by a quadratic form as shown below,
\begin{equation}
\label{V2}
\centering
V(\phi)=V_2(\kappa\phi)^2,
\end{equation}
where $V_2$ represents the amplitude of the inflaton's potential and has mass dimensions of eV$^4$. The dimensionless scalar coupling function $h(\phi)$ is assumed to have a power-law form,
\begin{equation}
\centering
\label{h2}
h(\phi)=1+\lambda(\kappa\phi)^2\, ,
\end{equation}
with $\lambda$ being a dimensionless parameter while the Chern-Simons scalar coupling function is given by a power-law form,
\begin{equation}
\centering
\label{v2}
\nu(\phi)=(\kappa\phi)^2\, .
\end{equation}
The quadratic coupling is chosen so that a blue-tilted tensor spectral index can be manifested. In addition, the slow-roll parameters of the model are described by simple forms, as shown below,
\begin{equation}
\centering
\label{e1model2}
\epsilon_1\simeq \frac{2-2 \lambda  \kappa ^2 \phi ^2}{\lambda  \kappa ^4 \phi ^4+\kappa ^2 \phi ^2}\, ,
\end{equation}

\begin{equation}
\centering
\label{e2model2}
\epsilon_2\simeq 2 \frac{2 \lambda  \left(6 \lambda ^2 \kappa ^4 \phi ^4+11 \lambda  \kappa ^2 \phi ^2+3\right)}{\lambda  \kappa ^2 \phi ^2+1}\, ,
\end{equation}

\begin{equation}
\centering
\label{e3model2}
\epsilon_3\simeq  \frac{2 \lambda  \left(\kappa ^2 \lambda  \phi ^2-1\right)}{\kappa ^2 \lambda  \phi ^2+1}\, ,
\end{equation}

\begin{equation}
\centering
\label{e4model2}
\epsilon_4\simeq \frac{2 \lambda  (6 \lambda +1) \left(\kappa ^2 \lambda  \phi ^2-1\right)}{\kappa ^2 \lambda  (6 \lambda +1) \phi ^2+1}\, ,
\end{equation}
except from index $\epsilon_5$, given that it carries the impact of the parity violation. It becomes clear that apart from the $\phi\to-\phi$ symmetry, parameter $\lambda$ which comes from the non-minimal part has a major impact on the inflationary indices. The aforementioned symmetry suggests that the initial and final value has either a positive of negative sign however, in order to obtain a viable phenomenology, only the positive sign shall be considered. In this case, the evolution of the scalar field implies that the scalar field decreases in magnitude as it evolves dynamically and when the vacuum expectation value of the scalar field is reached, the non-minimal part vanishes identically leaving us with $h(\phi_{vac})=1$. In other words, the effective Planck mass reaches its expected value. 
In order to proceed with the phenomenology of the model we set the first slow-roll index, as presented in Eq. (\ref{e1model2}) equal to unity. Then, the final value of the inflaton when the inflationary era ceases is,
\begin{equation}
\centering
\label{ffinal1}
\phi_{end}\simeq \frac{1}{\kappa} \sqrt{\frac{\sqrt{4 \lambda ^2+12 \lambda +1}-2 \lambda -1}{2\lambda  }}\, .
\end{equation}
Similarly, by making use of this value one can determine the relation which describes the inflaton during the first horizon crossing. Specifically by utilizing Eq. (\ref{efolds}) we end up with,
\begin{equation}
\centering
\label{finitial2}
\phi_k\simeq \frac{\sqrt{e^{-4 \lambda  N} \left(\kappa ^2 \lambda  \phi _{end}^2-1\right)+1}}{\kappa  \sqrt{\lambda }}\, .
\end{equation}
By designating the free parameters of the model in Natural units as $(\lambda,V_2,N)=(10^{-5},0.01M_P^4,60)$ one finds that the observational indices are in agreement with the latest Planck data as $n_\mathcal{S}=0.966941$ and $r=0.0210541$. The interesting feature of the model is that the expected value of the tensor spectral index is $n_\mathcal{T}=0.000369782$. In addition, the numerical values of the inflaton during the first horizon crossing and when the inflationary epoch ends are, $\phi_{end}=15.5469M_P$ and $\phi_{end}=1.41419M_P$ respectively which aligns with the fact that a decreasing evolution is expected. Concerning the slow-roll indices, their initial values are,
$\epsilon_1=0.0082346$, $\epsilon_2=0.0000604$, $\epsilon_3\simeq \epsilon_4 \simeq -0.0000199$ and finally, $\epsilon_5=-0.00841946$. Since $\epsilon_5<-\epsilon_1$ a blue spectrum is generated for the model at hand. Also, the fact that the slow-roll parameters obtain such small values is indicative of the fact that the slow-roll assumption is indeed respected.
However, we stress that the Swampland Criteria are not satisfied for this specific set of parameters. This can easily be inferred from the difference between $\phi_k$ and $\phi_{end}$ however, even the second criterion fails to be satisfied for the model at hand.

\begin{figure}[h!]
\centering
\includegraphics[width=15pc]{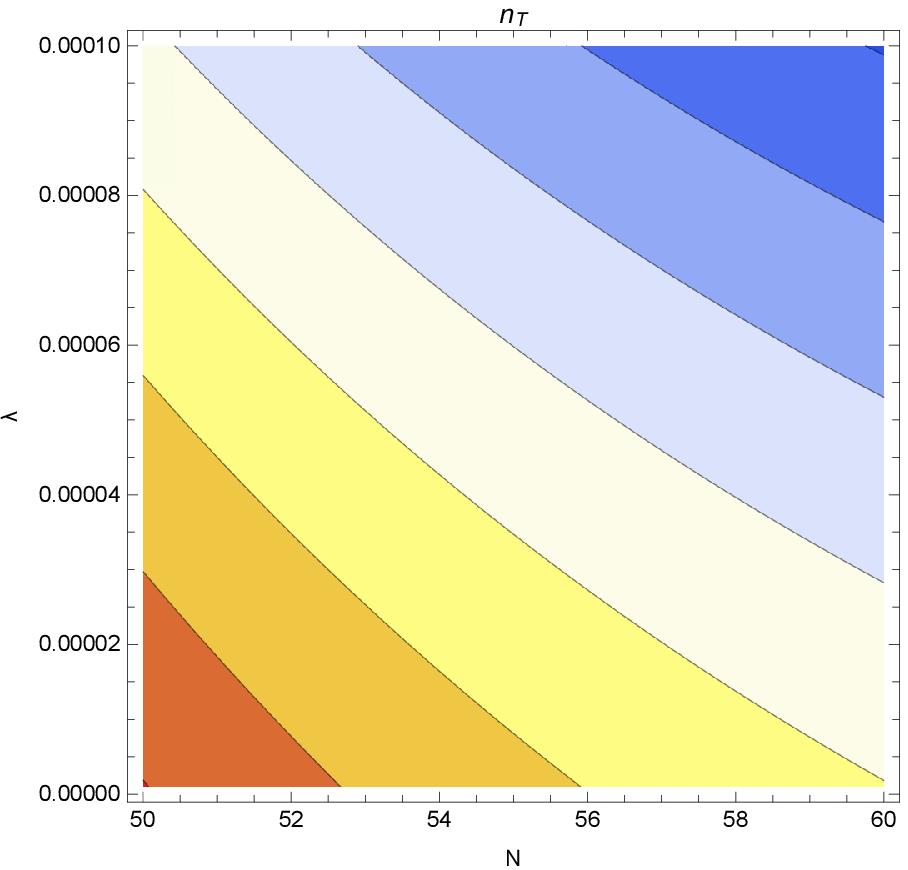}
\includegraphics[width=3pc]{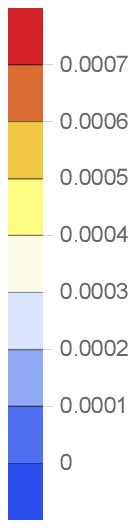}
\caption{Contour plot of the tensor spectral index $n_\mathcal{T}$ as a function of the e-folding number $N$ and auxiliary parameter $\lambda$. In this case it becomes clear that there exist a plethora of pairs of values that result in a blue spectrum.}
\label{ntmodelb}
\end{figure}

In Fig. \ref{ntmodelb} we present the behaviour of the tensor spectral index, which depends on the e-folding number and the free parameter $\lambda$. As it is obvious, accepted values for this index can be obtained for a wide range of parameters. Specifically, large values for $\lambda$ seem to result in a relative decrease of the tensor spectral index and around $\lambda\sim\mathcal{O}(10^{-3})$ viability of the scalar spectral index is spoiled. Furthermore, even though the Swampland criteria are not satisfied, the Lyth-bound in Eq. (\ref{Lythbound}) is respected since the inequality $7.7<14.1328$ is obtained in this scenario. Here, the field range seems to be quite large relative to the previous example.

\section{Constant-Roll Inflationary Phenomenology}
In this section we investigate an alternative scenario of the inflationary phenomenology of the non-minimally coupled Einstein-Chern-Simons gravity considering the constant-roll assumption. Specifically, the slow-roll approximations hold true during the inflationary era but, the scalar field varies with a constant rate of roll. Hence, our approximations are,
\begin{align}
\centering
\label{constantroll}
\dot H&\ll H^2\, ,&\frac{1}{2}\dot\phi^2&\ll V\, ,& \ddot\phi&= \beta H\dot\phi\, ,
\end{align}
where $\beta$ denotes the constant-roll parameter and for simplicity is considered to be $\phi$ independent, although this is not mandatory, see for instance  \cite{Oikonomou:2021yks}. The formalism that is used has already been presented in Sec.II thus, all that remains is to demonstrate the alterations between the two scenarios. Note that in Eq. (\ref{constantroll}), no statement was made about the non-minimal coupling since, conclusions can be derived from the constant-roll condition however, a viable inflationary phenomenology requires that $\ddot h\ll H\dot h$. This will become more clear in the following.

Under the constant-roll approximation
the system of equations of motion is,
\begin{equation}
\centering
\label{motion1constantroll}
H^2\simeq\frac{\kappa^2 V}{3h}\, ,
\end{equation}

\begin{equation}
\centering
\label{motion2constantroll}
\dot H\simeq-\frac{\kappa^2\dot\phi^2}{2h}\bigg(1+\frac{h''}{\kappa^2}\bigg)+\frac{H\dot h}{2h}(1-\beta)\, ,
\end{equation}

\begin{equation}
\centering
\label{motion3constantroll}
H\dot\phi(3+\beta)+V'-\frac{R}{2\kappa^2}h'\simeq0\, ,
\end{equation}
where the first time derivative of the scalar field can be extracted from the Eq. (\ref{motion3constantroll}) as before. In this approach the constant-roll parameter is involved in the time derivative of the scalar field as shown below,
\begin{equation}
\centering
\label{fdotconstantroll}
\dot\phi\simeq-\frac{V'-\frac{6H^2}{\kappa^2}h'}{H(3+\beta)}\, .
\end{equation}
Note that
the equation for the e-folds is also altered since, the first time derivative of the scalar field participates in the denominator. In the following, an inflationary model under the constant-roll assumption is presented. Hence, the constant-roll parameter now affects the numerical values for $\phi_k$ and $\phi_{end}$, so in other words it participates effectively in the overall phenomenology. Its numerical value however should not be considered to be arbitrarily large since, compatibility of the scalar spectral index with observations is spoiled.

\

\subsection{Quartic Chern-Simons Coupling}
As in the previous models, we commence our analysis by 
designating the scalar functions. Specifically,
the scalar potential
is considered to have a power-law form,
\begin{equation}
\label{V3}
\centering
V(\phi)=V_3(\kappa\phi)^2\, ,
\end{equation}
while, the rest scalar coupling functions are,
\begin{equation}
\label{h3}
\centering
h(\phi)=\gamma(\kappa\phi)^{-2}\, ,
\end{equation}
\begin{equation}
\label{v3}
\centering
\nu(\phi)=\frac{\kappa^4V(\phi)}{h(\phi)}\, ,
\end{equation}
where $\gamma$ is a dimensionless auxiliary parameter. In this approach, it becomes clear that the Chern-Simons scalar coupling function is also a power-law model with a relatively large exponent. Assuming the constant-roll paradigm, the slow-roll indices are given by the following relations,
\begin{equation}
\centering
\label{e1model3}
\epsilon_1 \simeq \frac{18 \gamma  \left(\left(\beta ^2+2 \beta +6\right) \kappa ^4 \phi ^4+54 \gamma \right)}{(\beta +3)^2 \kappa ^8 \phi ^8}\, ,
\end{equation}

\begin{equation}
\centering
\label{e2model3}
\epsilon_2 = \beta\, ,
\end{equation}

\begin{equation}
\centering
\label{e3model3}
\epsilon_3 \simeq \frac{18 \gamma }{(\beta +3) \kappa ^4 \phi ^4}\, ,
\end{equation}

\begin{equation}
\centering
\label{e4model3}
\epsilon_4 \simeq \frac{18 \gamma  \left(18 \gamma +\kappa ^4 \phi ^4\right)}{(\beta +3) \kappa ^4 \phi ^4 \left(6 \gamma +\kappa ^4 \phi ^4\right)}\, ,
\end{equation}
where index $\epsilon_5$ is once again omitted due its perplexed form. Here, it becomes clear that the constant-roll parameter affects the slow-roll indices however, in order to obtain a viable scalar spectral index, its value is assumed to be of order $\mathcal{O}(10^{-3})$. Following the same procedure as in the slow-roll case, it can easily be inferred that the values of the scalar field are,

\begin{widetext}
\begin{align}
\centering
\label{finalvaluemodel3}
\phi_{end}& \simeq \sqrt{3} \sqrt[4]{\frac{\left(\beta ^2+2 \beta +6\right) \gamma  \kappa ^4+\sqrt{\left(\beta ^4+4 \beta ^3+28 \beta ^2+96 \beta +144\right) \gamma ^2 \kappa ^8}}{(\beta +3)^2 \kappa ^8}}\, ,\\
\phi_k &\simeq \sqrt{3} \sqrt[4]{\frac{\sqrt{\left(\beta ^4+4 \beta ^3+28 \beta ^2+96 \beta +144\right) \gamma ^2 \kappa ^8}+\gamma  \kappa ^4 \left(\beta ^2+2 \beta +8 (\beta +3) N+6\right)}{(\beta +3)^2 \kappa ^8}}\, .
\end{align}
\end{widetext}

Proceeding to the numerical solutions, we set the following values for the free parameters, $(V_3,
\gamma,N,\beta)$=$(2\cdot10^{-4}M_P^4,2\cdot 10^{-6},64,0.0036)$. According to the aforementioned values the scalar spectral index of primordial perturbations and the tensor-to-scalar ratio are in agreement with the latest Planck data \cite{Planck:2018jri} as $n_\mathcal{S}=0.961374$ and $r=0.00673977$ respectively. The model also predicts a positive tensor spectral index as $n_\mathcal{T}=0.000248962$. Moreover, at the start of inflation the numerical value of the scalar field is $\phi_k=0.236042M_P$ and decreases during the inflationary epoch by reaching the value $\phi_{end}=0.0774365M_P$. The numerical results of the slow-roll parameters, that govern the cosmological dynamics during inflation are $\epsilon_1=0.00776675$, $\epsilon_3=0.003861, \epsilon_4=0.00389074$ and $\epsilon_5=-0.00789026$. These values imply that the slow-roll assumption imposed initially is indeed valid.

\begin{figure}[h!]
\centering
\includegraphics[width=15pc]{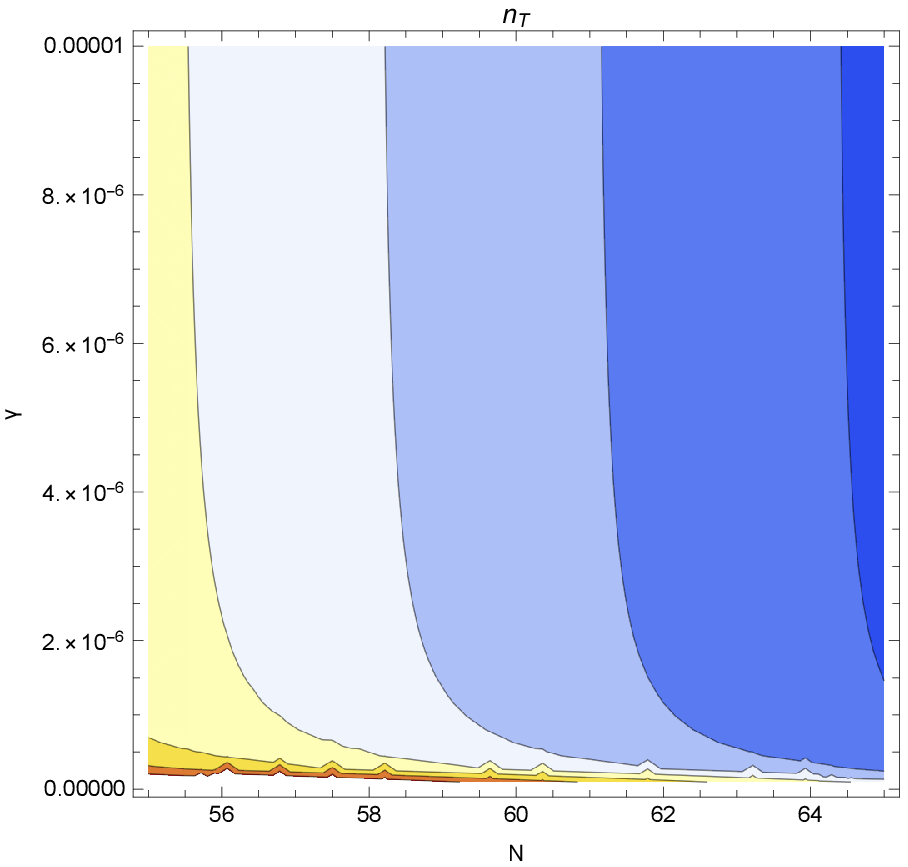}
\includegraphics[width=3pc]{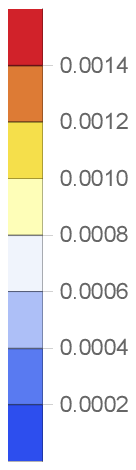}
\caption{Tensor spectral index $n_\mathcal{T}$ depending on e-folding number $N$ and $\gamma$. As shown, there are multiple values that generate a positive spectral index.}
\label{ntmodelc}
\end{figure}

At this point let us examine numerically whether the Swampland criteria are satisfied for this specific set of parameters. The Swampland distance conjecture, which involves the difference between the initial and final value of the inflaton is less than the unity in absolute value, hence the model can be treated as an effective model. The de Sitter conjecture is also satisfied since, 
$\frac{V'(\phi_k)}{\kappa V(\phi_k)}=8.47305$ is greater than unity. The supplementary criterion, as explained before, is not satisfied due to the fact that a power-law model is chosen.

\begin{figure}[h!]
\centering
\includegraphics[width=15pc]{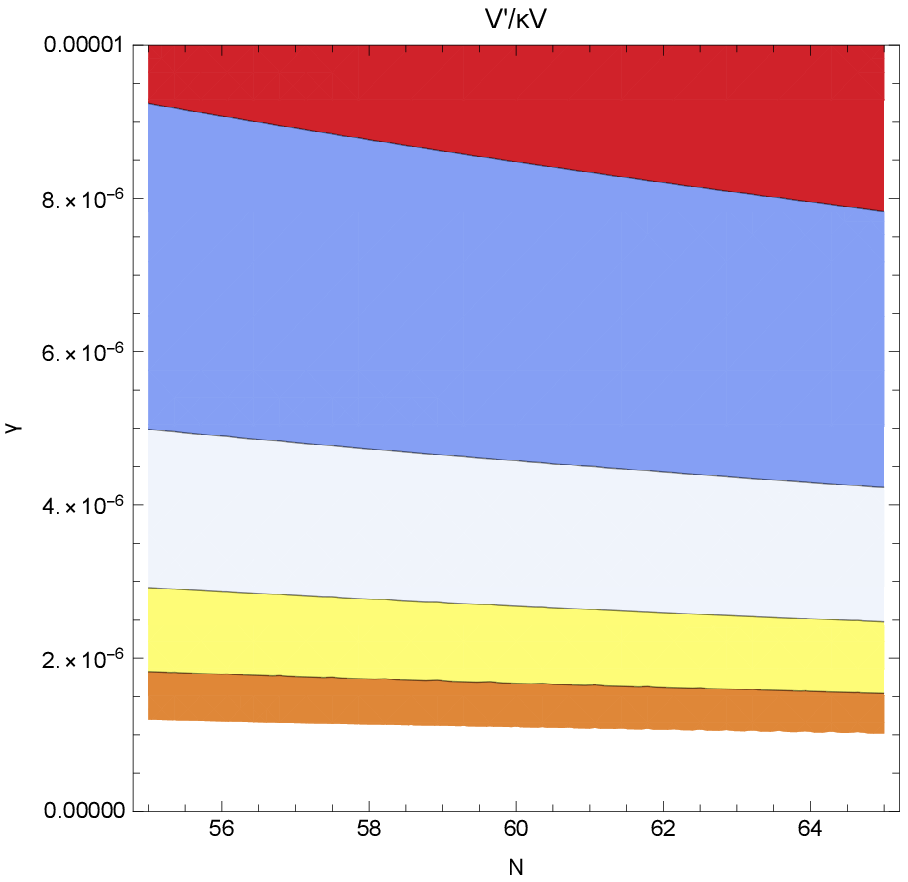}
\includegraphics[width=2pc]{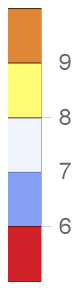}
\caption{Second criterion as a function of $N$ and $\gamma$. Here, larges values seem to increase the derivative of the scalar potential during the first horizon crossing.}
\label{Swamp3}
\end{figure}

Finally, Fig. \ref{ntmodelc} depicts the behaviour of the tensor spectral index of primordial perturbations for a wide set of values of the free parameters $N$ and $\gamma$. According to the density plot, higher values of the tensor spectral index can be manifested mainly for large e-folds. In addition, the ratio $\frac{V'}{\kappa V}$ at the start of inflation is presented in Fig.\ref{Swamp3} depending on the same parameters. Here, it becomes clear that small values of $\gamma$ which result in an effective increase of the value of the tensor spectral index, manage to also increase the derivative of the scalar potential.

 Concerning the Lyth-bound, under the constant-roll assumption, Eq. (\ref{Lythbound}) can be rewritten as,
 
\begin{equation}
\centering
\label{Lythboundconstantroll}
\bigg|\bigg(1+\frac{\beta}{3}\bigg)h\bigg(\frac{V'}{\kappa V}-2\frac{h'}{\kappa h}\bigg)\bigg|\Delta N<|\kappa\Delta\phi|\, ,
\end{equation}
and it can be shown that it is indeed satisfied while respecting the  Swampland criteria as $0.054813<0.158606<1$. Note that for $\beta\to0$, the previous result is extracted.

\section{Conclusions}

In this work the inflationary era of the primordial Universe was examined considering that the gravitational action involves a canonical scalar field and the Chern-Simons string corrective term. In addition, it was assumed that a dynamically evolving scalar function couples with the Ricci scalar. According to the formalism which was developed, all the involved physical quantities and their derivatives were expressed in terms of the scalar field namely, the inflaton. Several models were investigated with power-law potentials under the slow-roll and the constant-roll assumption. As shown, all of them were compatible with the latest Planck data by predicting the existence of a positive tensor spectral index due to the inclusion of the gravitational Chern-Simons term. This result can be connected to the amplification of the energy spectrum of primordial gravitational waves and may explain a possible future signal obtained in subsequent years. Finally, the Swampland criteria were investigated and as shown, non-minimally coupled models can in fact be regarded as effective descriptions under proper circumstances. The models at hand seem to also respect the Lyth-bound however due to the inclusion of the non-minimal coupling function and the satisfaction of the Swampland criteria, the bound differs from the canonical scalar field. Provided that a blue-tilted tensor spectral index is manifested, it would be interesting to numerically compute exponent $\mathcal{D}$ between the current era and high redshifts that belong to the radiation domination era. Afterwards, the energy spectrum of gravitational waves can be extracted due to the positive spectral index, the amplitude in the high frequency regime can be amplified. This amplification could be enough so that next generation detectors are able to measure it and at last shine light towards the mysterious early era of the Universe. This analysis can also be combined with additional scalar corrections from string inspired models or from $f(R)$ gravity. We hope to address this issue in future works.

\subsection*{Data Availability Statement}
No Data associated in the manuscript.

\end{document}